\documentclass[twocolumn]{aastex631}
\usepackage{amsmath,bm}
\usepackage{graphicx,subfigure}

\begin{document}

\title{Magnetic effect on the evolution of an eccentric and inclined orbit}

\author[0000-0002-3641-6732]{Xing Wei}
\affiliation{IFAA, School of Physics and Astronomy, Beijing Normal University, Beijing 100875, China}
\email{xingwei@bnu.edu.cn}

\begin{abstract}
We provide a method to calculate the evolution of an eccentric and inclined orbit under the magnetic effect. Taking the unipolar interaction as an example, we study both coplanar and inclined orbits. We calculate the Lorentz force and then the changes in orbital energy and angular momentum. Combined with the total energy dissipation we write down the evolution equations of semi-major axis, eccentricity, inclination and spin. This set of equations can be used to study the orbital and spin evolution under the magnetic interaction.
\end{abstract}

\section{Introduction} \label{introduction}

In addition to tidal interaction \citep{goldreich1966, hut1981, ogilvie2014}, magnetic interaction also plays an important role in orbital evolution of two-body system \citep{goldreich1969, laine2008, laine2012, lai2012}. Magnetic interaction has three types \citep{zarka2007, strugarek2018}: 1. the interaction of unmagnetized stellar wind and companion's magnetic field where companion can be either star or planet, 2. the interaction of stellar dipolar field and companion's dipolar field, and 3. the interaction primary's magnetic field and companion's orbital motion where primary can be star or planet and companion can be planet or moon. The second type is called dipole-dipole interaction and the third type unipolar interaction. Recently, \citet{strugarek2021} applied both tidal and magnetic interactions to study the orbital evolution of close-in planets, \citet{wei2024} studied the survival condition of exomoons, and \citet{owen2025} applied the combination of unipolar and tidal interactions to explain the observed patterns in the distribution of orbital periods and radii of rocky exoplanets.

However, all the studies about magnetic interaction consider only circular orbit, i.e., only the evolution of semi-major axis and spin is studied. How magnetic interaction influences eccentricity and inclination is the purpose of our work. \citet{hut1981} studied the tidal evolution of eccentric and inclined orbit, and we almost follow his method to study magnetic interaction with Lorentz force instead of tidal force. The difference is that we can also calculate total energy dissipation, i.e., ohmic dissipation, so that we can simplify our calculation. In section 2 the coplanar orbit is studied, in section 3 the inclined orbit is studied, and in section 4 some discussions are give.

\section{Coplanar orbit} \label{coplanar}

In this section we study the coplanar orbit with eccentricity, i.e., primary's spin axis is aligned with orbital angular momentum but orbit is eccentric. Suppose an elliptical orbit with semi-major axis $a$ and eccentricity $e$, and primary spins at angular velocity $\Omega$. The orbit in polar coordinates $(r,\theta)$ is described as
\begin{equation}\label{orbit}
r=\frac{a(1-e^2)}{(1+e\cos\theta)}
\end{equation}
where the origin is at primary's centre. In Cartesian coordinates $(x=r\cos\theta,y=r\sin\theta)$ companion's velocity relative to primary's spin reads
\begin{equation}\label{v}
\bm v=(\dot x+\Omega y)\hat{\bm x}+(\dot y-\Omega x)\hat{\bm y}.
\end{equation}
Electric current density reads $\bm J=\sigma\bm v\times\bm B$ where $\sigma$ is companion's electric conductivity and $\bm B=B\hat{\bm z}$ is primary's magnetic field at companion's location. Suppose that primary's field is dipolar and decays as $r^{-3}$, so that 
\begin{equation}\label{B}
\bm B=B_1(r/R_1)^{-3}
\hat{\bm z}
\end{equation}
where $B_1$ is magnetic field at primary's surface and $R_1$ is primary's radius. We assume that electric current is uniformly distributed in companion. Lorentz force then reads $\bm F=V_2\sigma(\bm v\times\bm B)\times\bm B$ where $V_2$ is companion's volume. Inserting the expressions of $\bm v$ \eqref{v} and $\bm B$ \eqref{B} and converting from Cartesian to polar coordinates, we find the radial and azimuthal components of $\bm F$
\begin{align}
& F_r=-V_2\sigma B_1^2(r/R_1)^{-6}\,\dot r, \label{force1} \\
& F_\theta=-V_2\sigma B_1^2(r/R_1)^{-6}\,r(\dot\theta-\Omega). \label{force2}
\end{align}

On the other hand, orbital energy and angular momentum read
\begin{align}
& E_o=-GM_1M_2/(2a), \label{Eo} \\
& h_o=\mu\sqrt{GMa(1-e^2)} \label{ho}
\end{align}
where $\mu=M_1M_2/(M_1+M_2)$ is reduced mass and $M=M_1+M_2$ is total mass. Compared to primary we neglect companion's size, so that total energy and angular momentum read
\begin{align}
& E=E_o+(1/2)I_1\Omega^2, \label{E} \\
& h=h_o+I_1\Omega \label{h}
\end{align}
where $I_1=\alpha M_1R_1^2$ is primary's moment of inertia and the coefficient $\alpha$ depends on density distribution.

Both the radial and azimuthal components of Lorentz force contribute to orbital energy $E_o$, but only the azimuthal component to orbital angular momentum $h_o$. The changes in $E_o$ and $h_o$ for one orbit read
\begin{align}
& \Delta E_o=\oint_{\rm orb} F_rdr+\oint_{\rm orb} F_\theta rd\theta = \int_0^{2\pi}\left(F_r\frac{dr}{d\theta}+F_\theta r\right)d\theta, \label{orbital-integral-1} \\
& \Delta h_o = \oint_{\rm orb} F_\theta rdt=\int_0^{2\pi} F_\theta\frac{\mu}{h_o}r^3d\theta \label{orbital-integral-2}
\end{align}
where the identities $dr=(dr/d\theta)d\theta$, $dt=d\theta/\dot\theta$ and $\dot\theta=h_o/(\mu r^2)$ are employed. Inserting the expressions of Lorentz force \eqref{force1}-\eqref{force2}, of $r$ \eqref{orbit} and of $dr/d\theta$ derived from \eqref{orbit} into \eqref{orbital-integral-1}-\eqref{orbital-integral-2} and employing the identities $\dot r=(dr/d\theta)\dot\theta$ and $\dot\theta=h_o/(\mu r^2)$, we integrate to obtain
\begin{align}
\Delta E_o=2\pi V_2\sigma B_1^2R_1^6 & \left[-(h_o/\mu)a^{-6}(1-e^2)^{-6}f_1\right.  \nonumber \\
& \left. +\Omega a^{-4}(1-e^2)^{-4}f_2\right], \label{DeltaEo} \\
\Delta h_o=2\pi V_2\sigma B_1^2R_1^6 & \left[-a^{-4}(1-e^2)^{-4}f_2\right. \nonumber \\
& \left. +(\mu/h_o)\Omega a^{-2}(1-e^2)^{-2}f_3\right] \label{Deltaho}
\end{align}
where the functions about eccentricity are
\begin{align}
f_1 & =1+8e^2+(51/8)e^4+(3/8)e^6,  \\
f_2 & =1+3e^2+(3/8)e^4,  \\
f_3 & =1+(1/2)e^2.
\end{align}

The Keplerian orbital period is 
\begin{equation}\label{P}
P=2\pi(GM/a^3)^{-1/2}.
\end{equation}
Substituting $\Delta E_o/P=\dot E_o$ and $\Delta h_o/P=\dot h_o$ into \eqref{Eo}-\eqref{ho}, we derive the migration and circularization timescales
\begin{align}
\dot a/a & =-(\Delta E_o/E_o)P^{-1} , \label{dota} \\
\dot e/e & =-\left[\Delta E_o/(2E_o)+\Delta h_o/h_o\right](1/e^2-1)P^{-1}. \label{dote}
\end{align}
The right-hand-side of \eqref{dota} and \eqref{dote} depends on $a$ and $e$ (see \eqref{Eo}-\eqref{ho}, \eqref{DeltaEo}-\eqref{Deltaho} and \eqref{P}), so that we can integrate \eqref{dota} and \eqref{dote} to obtain the evolution of $a$ and $e$.

Next, we calculate the spin evolution $\dot\Omega$. Of the binary system, the total angular momentum $h$ \eqref{h} conserves such that we derive the synchronization timescale
\begin{equation}\label{dotOmega}
\dot\Omega/\Omega=-\Delta h_o/(I_1\Omega)P^{-1}.
\end{equation}
Until now, we have derived the equations of $\dot a$, $\dot e$ and $\dot\Omega$. Given the initial orbital parameters $a$, $e$ and $\Omega$, we can integrate \eqref{dota}, \eqref{dote} and \eqref{dotOmega} to achieve the evolution of $a$, $e$ and $\Omega$.

For a circular orbit $e=0$ around a non-rotating primary $\Omega=0$, the magnetic torque $\Delta h_o/P$ reduces to scale as $a^{-11/2}$, consistent with a straightforward calculation where magnetic torque scales as $\propto (a/P)B^2a\propto a^{-11/2}$. Compared to tidal torque $\propto a^{-6}$, magnetic torque changes with distance less sharper than tidal torque, implying that it dominates over tidal torque at a relative long separation \citep{wei2024}.

To end this section, we calculate the total energy dissipation rate by \eqref{E}
\begin{equation}\label{energy-dissipation}
\dot E=-E_o(\dot a/a)+I_1\Omega\dot\Omega=\left(\Delta E_o-\Omega\Delta h_o\right)P^{-1}.
\end{equation}
Inserting \eqref{DeltaEo}-\eqref{Deltaho} into \eqref{energy-dissipation} we can obtain the total energy dissipation rate. As we know, the instantaneous energy dissipation arises from the ohmic dissipation $-V_2J^2/\sigma$ where the electric current density is $\bm J=\sigma\bm v\times\bm B$. Thus, the orbit-averaged dissipation rate reads
\begin{align}\label{ohmic-dissipation}
\dot E & =-V_2\sigma\oint_{\rm obt}B^2(v_x^2+v_y^2)dt\;P^{-1}  \nonumber \\
& =-\frac{V_2\sigma B_1^2R_1^6}{P}\int_0^{2\pi}\frac{\mu}{h_o}r^{-4}\left[\dot r^2+(r\dot\theta-\Omega r)^2\right]d\theta
\end{align}
where \eqref{v}, \eqref{B}, $dt=d\theta/\dot\theta$ and $\dot\theta=h_o/(\mu r^2)$ are employed. It is readily verified that \eqref{ohmic-dissipation} is identical to \eqref{energy-dissipation} by substituting \eqref{force1}-\eqref{force2} and \eqref{orbital-integral-1}-\eqref{orbital-integral-2} into \eqref{energy-dissipation} and employing $dr/d\theta=\mu r^2\dot r/h_o$.

\section{Inclined orbit} \label{inclined}

Recently, observations using Rossiter–McLaughlin effect discovered that hot stars with hot Jupiters have high obliquity \citep{winn2010}. In addition to tidal interaction, magnetic interaction may play a role in obliquity evolution. We go on studying the inclined orbit. To make our calculations convenient, we choose the orbital plane as the $x-y$ plane and set the origin at primary's centre, such that primary's spin and magnetic field are inclined to the $x-y$ plane. Although the orbital plane is non-inertial frame, the changes in orbital energy and angular momentum are not necessarily calculated in an inertial frame provided that the precession is fast. Denote the inclination angle by $\psi$ and the precession frequency of spin and magnetic field by $\omega$, and we write spin vector and magnetic dipole moment in Cartesian coordinates $(x,y,z)$ as
\begin{align}\label{Omega-m}
\bm\Omega & =\Omega(\sin\psi\cos\omega t,\sin\psi\sin\omega t,\cos\psi),  \\
\bm m & =m(\sin\psi\cos\omega t,\sin\psi\sin\omega t,\cos\psi).
\end{align}
Usually spin precession is much faster than orbital evolution. Magnetic field is calculated via magnetic dipole moment
\begin{equation}
\bm B=\mu/(4\pi)\left[3\bm r(\bm m\cdot\bm r)/r^5-\bm m/r^3\right]
\end{equation}
where $\bm r=(x,y,z)$ is position vector in 3D space and $\mu$ is magnetic permeability. The magnitude of magnetic dipole moment can be replaced with primary's surface field $m=4\pi B_1R_1^3/\mu$. Therefore, the magnetic field at companion $(x,y,z=0)$ reads
\begin{align}
B_x & =B_1R_1^3\sin\psi\frac{(2x^2-y^2)\cos\omega t+3xy\sin\omega t}{r^5}, \label{B3d1} \\
B_y & =B_1R_1^3\sin\psi\frac{(2y^2-x^2)\sin\omega t+3xy\cos\omega t}{r^5}, \label{B3d2} \\
B_z & =-B_1R_1^3\frac{\cos\psi}{r^3}. \label{B3d3}
\end{align}
On the other hand, the companion's velocity relative to primary's spin at companion $(x,y,z=0)$ reads
\begin{align}
v_x & =(\dot x+\Omega y\cos\psi), \label{v3d1} \\
v_y & =(\dot y-\Omega x\cos\psi), \label{v3d2} \\
v_z & =\Omega\sin\psi(x\sin\omega t-y\cos\omega t). \label{v3d3}
\end{align}
The Lorentz force $\bm F=V_2\sigma(\bm v\times\bm B)\times\bm B$ reads
\begin{align}
F_x & =V_2\sigma\left[B_x(v_yB_y+v_zB_z)-v_x(B_y^2+B_z^2)\right], \label{force3d1} \\
F_y & =V_2\sigma\left[B_y(v_xB_x+v_zB_z)-v_y(B_x^2+B_z^2)\right], \label{force3d2} \\
F_z & =V_2\sigma\left[B_z(v_xB_x+v_yB_y)-v_z(B_x^2+B_y^2)\right]. \label{force3d3}
\end{align}
Inserting \eqref{B3d1}-\eqref{B3d3} and \eqref{v3d1}-\eqref{v3d3} into \eqref{force3d1}-\eqref{force3d3}, we can write the Lorentz force at companion $(x,y,z=0)$ in terms of $x$, $y$, $\dot x$, $\dot y$, $\Omega$, $\psi$ and $\omega$. Then we average the Lorentz force over the fast precession, i.e., $\langle\cos^2\omega t\rangle=1/2$, $\langle\sin^2\omega t\rangle=1/2$, $\langle\sin\omega t\cos\omega t\rangle=0$, and the averaged first-order and third-order terms are 0. Thus, we are led to 
\begin{align}
F_x & = \frac{V_2\sigma B_1^2R_1^6}{2r^6} \left[ 3\cos\theta\sin\theta(\dot y-\Omega x\cos\psi)\sin^2\psi \right. \nonumber \\
& -\Omega y\sin^2\psi\cos\psi - (1+3\sin^2\theta)(\dot x+\Omega y\cos\psi)\sin^2\psi \nonumber \\
& \left. -2(\dot x+\Omega y\cos\psi)\cos^2\psi \right] , \label{averaged-force-1} \\
F_y & = \frac{V_2\sigma B_1^2R_1^6}{2r^6} \left[ 3\cos\theta\sin\theta(\dot x+\Omega y\cos\psi)\sin^2\psi \right. \nonumber \\
& +\Omega x\sin^2\psi\cos\psi - (1+3\cos^2\theta)(\dot y-\Omega x\cos\psi)\sin^2\psi \nonumber \\
& \left. -2(\dot y-\Omega x\cos\psi)\cos^2\psi \right], \label{averaged-force-2} \\
F_z & = 0. \label{averaged-force-3}
\end{align}

Now we follow the procedure in the last section to derive $\Delta E_o$ and $\Delta h_o$. Firstly, we convert the Lorentz force from Cartesian to cylindrical coordinates by these transformations: $x=r\cos\theta$, $y=r\sin\theta$, $\dot x=\dot r\cos\theta-r\dot\theta\sin\theta$, $\dot y=\dot r\sin\theta+r\dot\theta\cos\theta$, $F_r=F_x\cos\theta+F_y\sin\theta$ and $F_\theta=-F_x\sin\theta+F_y\cos\theta$,
\begin{align}
F_r &= \frac{V_2\sigma B_1^2R_1^6}{2r^6} \dot r \left[ \left(\frac{3}{4}\sin^2 2\theta-1\right)\sin^2\psi-2\cos^2\psi \right],  \\
F_\theta &= \frac{V_2\sigma B_1^2R_1^6}{2r^6} \left[ 2r\dot\theta(\cos^2\psi-2)+\Omega r\cos\psi(5-3\cos^2\psi) \right].
\end{align}
When $\psi=0$ the above expressions of $F_r$ and $F_\theta$ reduce to the coplanar case \eqref{force1}-\eqref{force2}. Secondly, with the expressions of $F_r$ and $F_\theta$, we calculate $\Delta E_o$ and $\Delta h_o$ by the orbital integrals \eqref{orbital-integral-1}-\eqref{orbital-integral-2},
\begin{align}
& \Delta E_o = \pi V_2\sigma B_1^2R_1^6 \left\{ (h_o/\mu)a^{-6}(1-e^2)^{-6}\times \right. \nonumber \\
& \left[ 2g_1(\cos^2\psi-2)+e^2\left(\left(\frac{3}{16}g_2-\frac{1}{2}g_3\right)\sin^2\psi-g_3\cos^2\psi\right) \right] \nonumber \\
& \left. +\Omega a^{-4}(1-e^2)^{-4}f_2\cos\psi(5-3\cos^2\psi) \right\}, \label{DeltaEo3d} \\
& \Delta h_o = \pi V_2\sigma B_1^2R_1^6 \left[ 2a^{-4}(1-e^2)^{-4}f_2(\cos^2\psi-2) \right. \nonumber \\
& \left. +(\mu/h_o)\Omega a^{-2}(1-e^2)^{-2}f_3\cos\psi(5-3\cos^2\psi) \right] \label{Deltaho3d}
\end{align}
where the functions about eccentricity are
\begin{align}
g_1 & = 1+(15/2)e^2+(45/8)e^4+(5/16)e^6,  \\
g_2 & = 1+(9/4)e^2+(3/16)e^4, \\
g_3 & = 1+(3/2)e^2+(1/8)e^4.
\end{align}
The expressions of $\Delta E_o$ \eqref{DeltaEo3d} and $\Delta h_o$ \eqref{Deltaho3d} for inclined orbit are more complex than \eqref{DeltaEo}-\eqref{Deltaho} for coplanar orbit. When $\psi=0$, \eqref{DeltaEo3d}-\eqref{Deltaho3d} reduce to \eqref{DeltaEo}-\eqref{Deltaho}.

Since the orbital plane is the $x-y$ plane, the equations of $\dot a$ and $\dot e$ are the same as \eqref{dota} and \eqref{dote}, just inserting the new $\Delta E_o$ and $\Delta h_o$. To derive the equations about $\dot\Omega$ and $\dot\psi$, we consider the conservation of total angular momentum
\begin{align}
& h_z=h_o+I_1\Omega\cos\psi, \label{total-h-z} \\
& h_{\perp}=0 \label{total-h-hori}
\end{align}
where we have averaged the fast precession. The $z$ component \eqref{total-h-z} yields
\begin{equation}\label{dotOmega3d}
(\dot\Omega/\Omega)\cos\psi-\dot\psi\sin\psi=-\Delta h_o/(I_1\Omega)P^{-1}.
\end{equation}
Compared to \eqref{dotOmega} of coplanar orbit, the term about $\dot\psi$ appears. The horizontal component \eqref{total-h-hori} at companion $(x,y,z=0)$ yields $F_z=0$, consistent with \eqref{averaged-force-3}.

Until now we have three equations \eqref{dota}, \eqref{dote} and \eqref{dotOmega3d} but four variables $a$, $e$, $\Omega$ and $\psi$. We use the equation of the orbit-averaged energy dissipation rate to close the system. According to the instantaneous energy dissipation rate $-V_2J^2/\sigma$ and $\bm J=\sigma\bm v\times\bm B$, we derive the orbit-averaged energy dissipation rate
\begin{align}\label{dotE}
\dot E & =-\frac{V_2\sigma}{P}\oint_{\rm orb}\left[v^2B^2-(v_xB_x+v_yB_y+v_zB_z)^2\right]dt \nonumber \\
& =-\frac{V_2\sigma B_1^2R_1^6}{P}\int_0^{2\pi}\frac{\mu}{h_o}r^2d\theta\times \nonumber \\
& \left\{\frac{\cos^2\psi}{r^6}A+\frac{\Omega^2\sin^2\psi\cos^2\psi}{2r^4}+\frac{5\sin^2\psi}{2r^6}A+\frac{7\Omega^2\sin^4\psi}{8r^4}\right. \nonumber \\
& \left. -\frac{1}{2}\left[(a_1+a_2+a_3)^2+(b_1+b_2+b_3)^2\right]\right\}
\end{align}
where the coefficients are
\begin{equation}
A={\dot r}^2+r^2{\dot\theta}^2+r^2\Omega^2\cos^2\psi-2r^2\dot\theta\Omega\cos\psi, 
\end{equation}
\begin{equation}
a_1=(3\cos^2\theta-1)(\dot x+\Omega y\cos\psi)\sin\psi/r^3,  
\end{equation}
\begin{equation}
a_2=3\cos\theta\sin\theta(\dot y-\Omega x\cos\psi)\sin\psi/r^3,  
\end{equation}
\begin{equation}
a_3=\Omega\sin\theta\cos\psi\sin\psi/r^2,  
\end{equation}
\begin{equation}
b_1=3\cos\theta\sin\theta(\dot x+\Omega y\cos\psi)\sin\psi/r^3,  
\end{equation}
\begin{equation}
b_2=(3\cos^2\theta-1)(\dot y-\Omega x\cos\psi)\sin\psi/r^3,  
\end{equation}
\begin{equation}
b_3=-\Omega\cos\theta\cos\psi\sin\psi/r^2.
\end{equation}
In the derivation of \eqref{dotE}, Equations \eqref{B3d1}-\eqref{B3d3}, \eqref{v3d1}-\eqref{v3d3}, $dt=d\theta/\dot\theta$ and $\dot\theta=h_o/(\mu r^2)$ are employed and precession has been averaged. The explicit expression of $\dot E$ is too complex and we have to use the numerical integration to calculate its value. According to \eqref{E} we can write 
\begin{equation}\label{close}
-(\dot a/a)E_o+I_1\Omega\,\dot\Omega=\dot E
\end{equation}
where $\dot E$ is calculated by \eqref{dotE}.

Now we can integrate \eqref{dota}, \eqref{dote}, \eqref{dotOmega3d} and \eqref{close} with the known $E_o(a)$, $h_o(a,e)$, $P(a)$, $\Delta E_o(a,e,\Omega,\psi)$, $\Delta h_o(a,e,\Omega,\psi)$ and $\dot E(a,e,\Omega,\psi)$ at each time step to obtain the evolution of $a$, $e$, $\Omega$ and $\psi$.

\section{Discussions}

In this short paper we derive the orbital evolution equations of a two-body system under the magnetic interaction with respect to coplanar and inclined orbits. What we consider here is only unipolar interaction so that the pre-factor of $\Delta E_o$ and $\Delta h_o$ is $V_2\sigma B_1^2 R^6$. When the circuit resistance is very small, the circuit can break due to the twist of magnetic flux tube and magnetic torque and dissipation rate can largely reduce \citep{lai2012}. With the estimated upper limit of torque and dissipation rate, \citet{lai2012} suggested that unipolar interaction is unimportant for compact objects, e.g., neutron star binaries and white dwarf binaries, but may be important for star-planet system at T Tauri phase. In this paper we provide a calculation method. Even if unipolar interaction is weak, some other magnetic interactions may work provided that the resistance is not too small. If we consider another magnetic interaction (say, dipole-dipole interaction) then what we need is only to replace the pre-factor $V_2\sigma B_1^2R_1^6$ with another one (say, a pre-factor related to Alfven Mach number \citep{zarka2007, li2025}) but the rest of equations keeps unchanged.

As we have already known, for a circular orbit the magnetic torque around a non-rotating primary scales as $a^{-11/2}$ whereas the tidal torque as $a^{-6}$. For an eccentric and inclined orbit whether the magnetic interaction is less sharper than tidal interaction needs to be numerical validated in the future study.

\section*{Acknowledgements}
I thank Zhoujian Cao for his discussion and Changzhi Lu for his check on my calculations.

\bibliography{paper}{}

\begin{thebibliography}{}
\expandafter\ifx\csname natexlab\endcsname\relax\def\natexlab#1{#1}\fi
\providecommand{\url}[1]{\href{#1}{#1}}
\providecommand{\dodoi}[1]{doi:~\href{http://doi.org/#1}{\nolinkurl{#1}}}
\providecommand{\doeprint}[1]{\href{http://ascl.net/#1}{\nolinkurl{http://ascl.net/#1}}}
\providecommand{\doarXiv}[1]{\href{https://arxiv.org/abs/#1}{\nolinkurl{https://arxiv.org/abs/#1}}}

\bibitem[{{Ahuir} {et~al.}(2021){Ahuir}, {Strugarek}, {Brun}, \&
  {Mathis}}]{strugarek2021}
{Ahuir}, J., {Strugarek}, A., {Brun}, A.~S., \& {Mathis}, S. 2021, Astron.
  Astrophys., 650, A126

\bibitem[{{De Colle} {et~al.}(2025){De Colle}, {Lin}, {Chen}, \& {Li}}]{li2025}
{De Colle}, F., {Lin}, D. N.~C., {Chen}, C., \& {Li}, G. 2025, \mnras, 539,
  3459

\bibitem[{{Goldreich} \& {Lynden-Bell}(1969)}]{goldreich1969}
{Goldreich}, P., \& {Lynden-Bell}, D. 1969, Astrophys. J., 156, 59

\bibitem[{{Goldreich} \& {Soter}(1966)}]{goldreich1966}
{Goldreich}, P., \& {Soter}, S. 1966, Icarus, 5, 375

\bibitem[{{Hut}(1981)}]{hut1981}
{Hut}, P. 1981, \aap, 99, 126

\bibitem[{{Lai}(2012)}]{lai2012}
{Lai}, D. 2012, Astrophys. J. Lett., 757, L3

\bibitem[{{Laine} \& {Lin}(2012)}]{laine2012}
{Laine}, R.~O., \& {Lin}, D. N.~C. 2012, Astrophys. J., 745, 2

\bibitem[{{Laine} {et~al.}(2008){Laine}, {Lin}, \& {Dong}}]{laine2008}
{Laine}, R.~O., {Lin}, D. N.~C., \& {Dong}, S. 2008, Astrophys. J., 685, 521

\bibitem[{{Lee} \& {Owen}(2025)}]{owen2025}
{Lee}, E.~J., \& {Owen}, J.~E. 2025, \apjl, 980, L40

\bibitem[{{Ogilvie}(2014)}]{ogilvie2014}
{Ogilvie}, G.~I. 2014, Annu. Rev. Astron. Astrophys., 52, 171

\bibitem[{{Strugarek}(2018)}]{strugarek2018}
{Strugarek}, A. 2018, in Handbook of Exoplanets, ed. H.~J. {Deeg} \& J.~A.
  {Belmonte} (Springer), 25

\bibitem[{{Wei} \& {Lin}(2024)}]{wei2024}
{Wei}, X., \& {Lin}, D.~N.~C. 2024, \apj, 965, 88

\bibitem[{{Winn} {et~al.}(2010){Winn}, {Fabrycky}, {Albrecht}, \&
  {Johnson}}]{winn2010}
{Winn}, J.~N., {Fabrycky}, D., {Albrecht}, S., \& {Johnson}, J.~A. 2010,
  Astrophys. J. Lett., 718, L145

\bibitem[{{Zarka}(2007)}]{zarka2007}
{Zarka}, P. 2007, Planet. Space Sci., 55, 598

\end{thebibliography}
\bibliographystyle{aasjournal}

\end{document}